\documentclass{article}

\def\xxinput#1{\input#1}

\xxinput{vsolj02.sty}

\usepackage[dvipdfmx]{graphicx}
\usepackage[comma,colon]{natbib}

\usepackage[OT2,T1]{fontenc}

\def\cite{\citealt}
\setcitestyle{aysep={}}

\newcounter{author}
\def\altaffilmark#1{$^{#1}$}
\def\altaffiltext#1{$^{#1}$\,}

\def\authorcount#1#2{{\refstepcounter{author}\label{#1}
                     \altaffiltext{\ref{#1}}{#2}}}

\begin{document}

\begin{center}

\title{V476 Cyg (Nova Cyg 1920) is currently a dwarf nova ---}
\vskip -2mm
\title{first such an object in the period gap?}

\author{
        Taichi~Kato\altaffilmark{\ref{affil:Kyoto}}
}
\email{tkato@kusastro.kyoto-u.ac.jp}

\authorcount{affil:Kyoto}{
     Department of Astronomy, Kyoto University, Sakyo-ku,
     Kyoto 606-8502, Japan}

\end{center}

\begin{abstract}
\xxinput{abst.inc}
\end{abstract}

\section{Introduction}

   V476 Cyg was discovered as a bright Galactic nova
by \citet{den20v476cyg}.  The visual peak magnitude by
\citet{den20v476cyg} was 2.2 on 1920 August 23.
Photographic observations showed relatively slow
rise from 7.0~mag to the peak (2.0~mag) which took place
in 7~d \citep{cam32v476cygv603aql}.
This nova was a fast nova with $t_2$=16.5~d
\citep{due87novaatlas} or $t_2$=6~d \citep{str10novalc}.
The nova was classified as a D-class one with a weak
dust dip in the light curve by \citet{str10novalc}.
Leslie Peltier described that the post-nova could be
sometimes glimpsed in his autobiographical
\textit{Starlight Nights} \citep{pel65StarlightNights}\footnote{
   I read this story in the book translated to Japanese
   \citep{pel85StarlightNightsJP}.
}.
He indeed followed this nova since its maximum in 1920
(when he was at an age of 20) and
saw it around 16~mag or slightly below it between 1961 and
1972 according to the AAVSO International Database\footnote{
   $<$http://www.aavso.org/data-download$>$.
}.  \citet{rin96oldnovaspec} reported a low-resolution
spectrum at $V$=17.33.  \citet{Hibernation} reported
a magnitude of $V\sim$18.7 and \citet{rin96oldnovaspec}
suspected that either there were significant flux errors
in \citet{Hibernation} or the nova was variable
on a short time-scale.  \citet{rin96oldnovaspec}
discussed that wiggles in the spectrum of V476 Cyg
might be a signature of a dwarf nova.

\section{V476 Cyg as a dwarf nova}

\begin{figure*}
\begin{center}
\includegraphics[width=16cm]{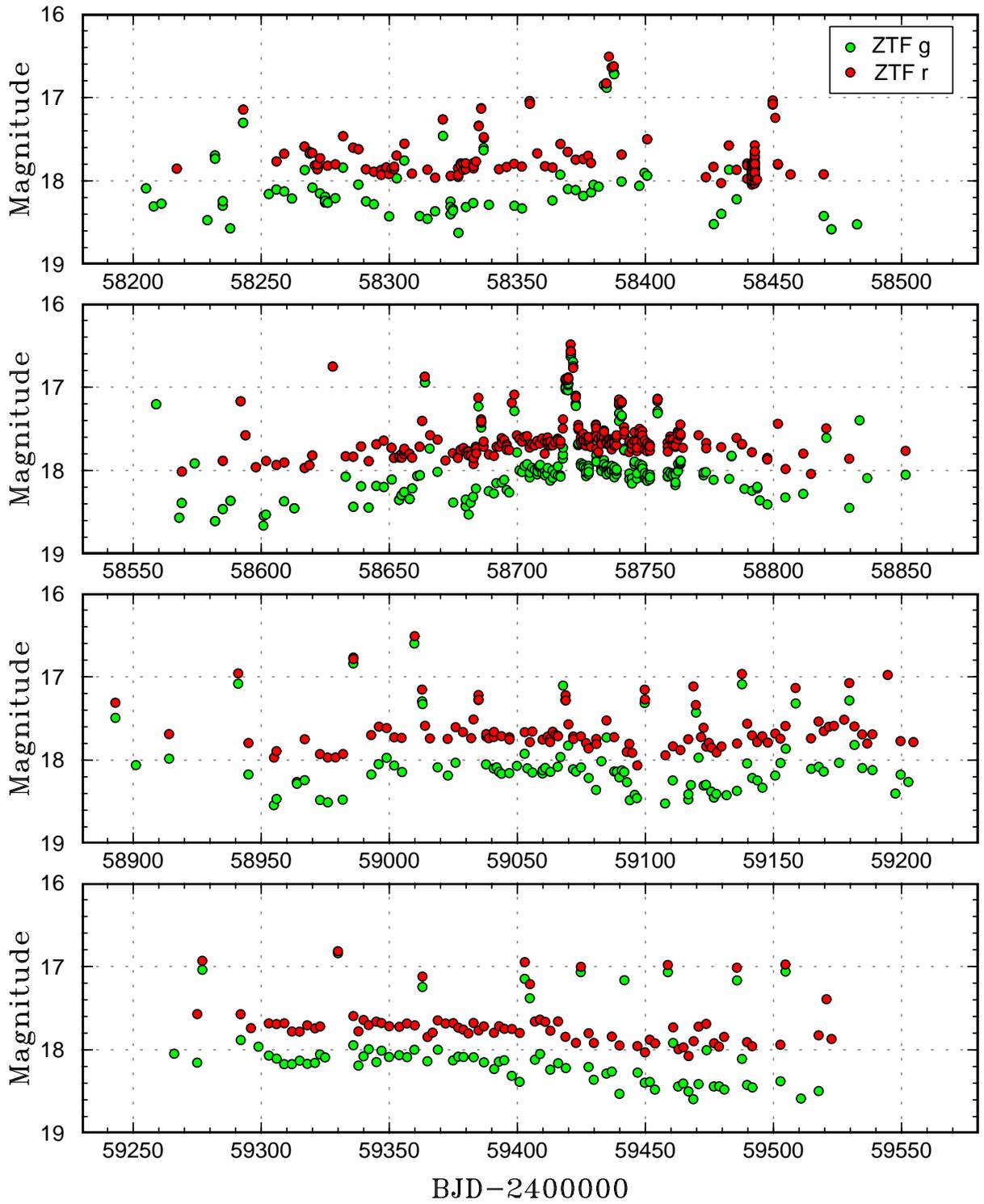}
\caption{
  ZTF light curve of V476 Cyg.
}
\label{fig:v476cyglc}
\end{center}
\end{figure*}

\begin{figure*}
\begin{center}
\includegraphics[width=15cm]{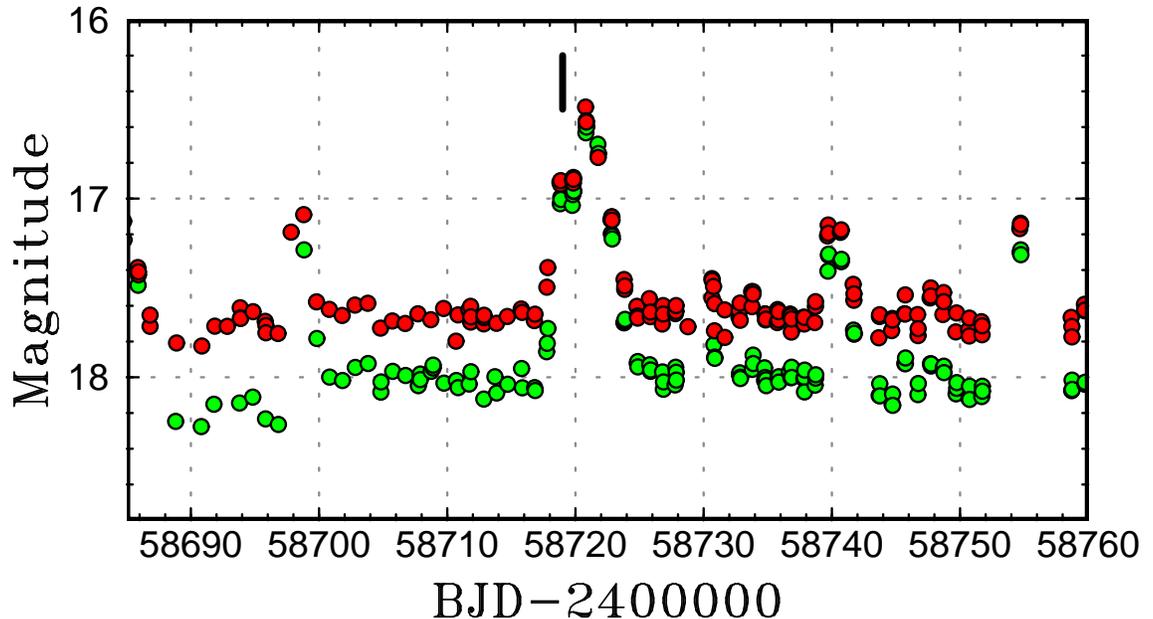}
\caption{
  ZTF light curve of V476 Cyg.  Enlargement of
  the bright state in 2019 August--September.
  The tick represents a shoulder in the bright
  outburst.  The symbols are the same as in
  figure \ref{fig:v476cyglc}.
}
\label{fig:v476cyglarge}
\end{center}
\end{figure*}

   Using the Zwicky Transient Facility (ZTF: \cite{ZTF})
public data\footnote{
   The ZTF data can be obtained from IRSA
$<$https://irsa.ipac.caltech.edu/Missions/ztf.html$>$
using the interface
$<$https://irsa.ipac.caltech.edu/docs/program\_interface/ztf\_api.html$>$
or using a wrapper of the above IRSA API
$<$https://github.com/MickaelRigault/ztfquery$>$.
}, I found that this object is currently
a dwarf nova (T. Kato on 2020 March 5, vsnet-chat 8457\footnote{
  $<$http://ooruri.kusastro.kyoto-u.ac.jp/mailarchive/vsnet-chat/8457$>$.
}) [for general information of cataclysmic variables and dwarf novae,
see e.g. \citet{war95book}].
Here I report on this object using the ZTF data
up to the end of 2021.
The light curve is shown in figure \ref{fig:v476cyglc}.
I must note, however, neither all outbursts were detected
nor all outbursts were detected at their peaks
by ZTF.
The quiescent brightness varied relatively strongly,
and the object was bright in 2019 August--September
(BJD 2458700--2458760).  During this bright phase, there
was a outburst starting on BJD 2458718 (2019 August 22;
figure \ref{fig:v476cyglarge}), which had a shoulder
[or referred to as an embedded precursor by 
\citet{can12ugemLC}] and the peak brightness
($r\sim$16.5 and $g\sim$16.6) was brighter than
the other outbursts.  There were equally bright outbursts
in 2018 September, peaking on BJD 2458386 (first panel
of figure \ref{fig:v476cyglc}) and in 2020 June, peaking
on BJD 2459010 (third panel of \ref{fig:v476cyglc}).
The former outburst apparently had a shoulder as in
the 2019 August one.

   The color was $g-r$=$+$0.1 at outburst peak, while
it was redder ($g-r\sim +$0.5) in quiescence.
This was probably due to the presence of a close,
physically unrelated, companion star
Gaia EDR3 2089624258068065152 with a Gaia magnitude
$G$=19.07 \citep{GaiaEDR3}.

\begin{figure*}
\begin{center}
\includegraphics[width=16cm]{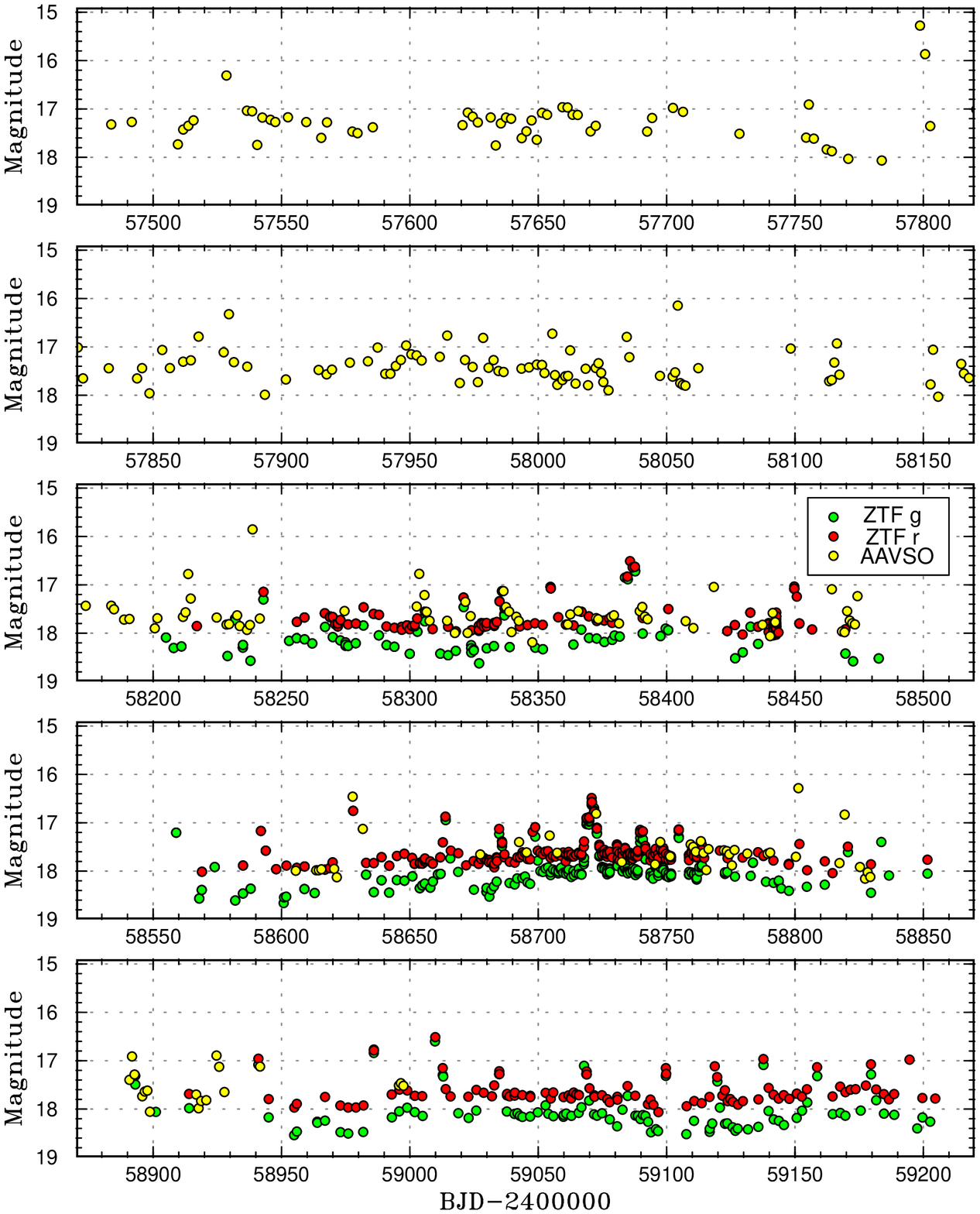}
\caption{
  Combined ZTF and AAVSO light curve of V476 Cyg.
  The AAVSO observations were unfiltered CCD ones
  obtained by HKEB (K. Hills, UK).  At least a few
  outbursts recorded by ZTF were also recorded by
  AAVSO CCD observations.
}
\label{fig:v476cyglcaavso}
\end{center}
\end{figure*}

   There were also CCD observations in
the AAVSO International Database between 2016 and 2019.
Short outbursts can be recognized by comparing with
the ZTF data (figure \ref{fig:v476cyglcaavso}).
The AAVSO observations were unfiltered CCD ones
obtained by HKEB (K. Hills, UK).  At least a few
outbursts recorded by ZTF were also recorded by
AAVSO CCD observations.
There was a bright outburst on
2017 February 14 (unfiltered CCD magnitude 15.3).
The dwarf nova state should have started before 2016.
Although there were some CCD observations with significant
variations in 2007, the data were not sufficient
to identify them as dwarf nova outbursts.

\section{Nova in the period gap?}\label{sec:porb}

\begin{figure*}
\begin{center}
\includegraphics[width=15cm]{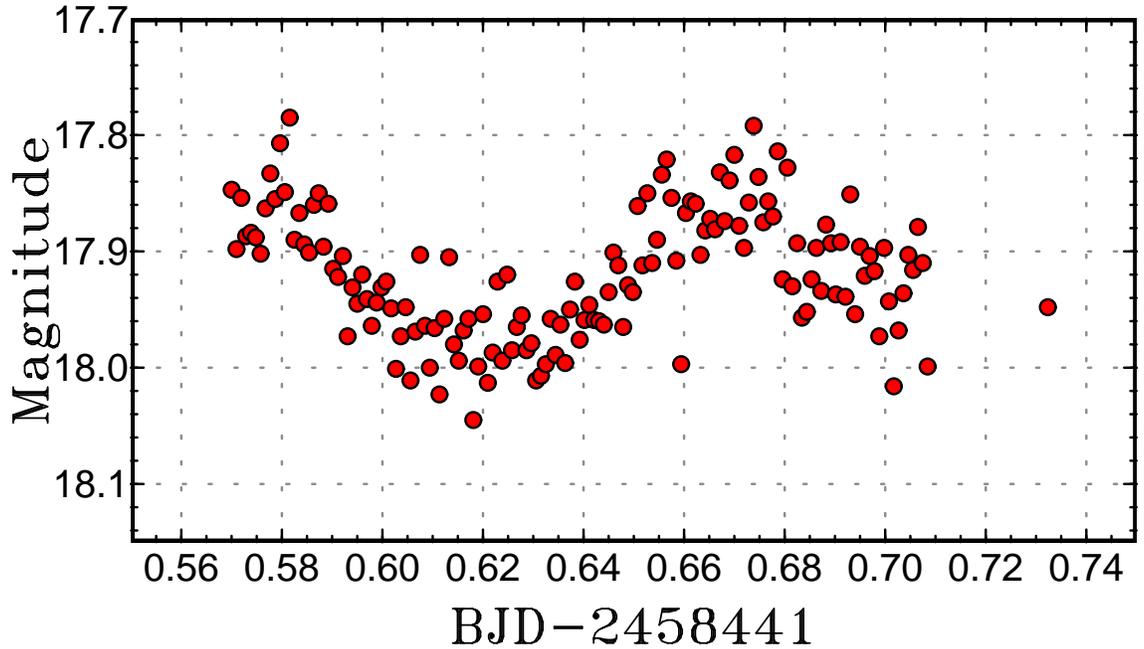}
\caption{
  Short-term variation recorded in $r$-band
  time-resolved photometry by ZTF.
}
\label{fig:v476cygshort}
\end{center}
\end{figure*}

\begin{figure*}
  \begin{center}
    \includegraphics[width=16cm]{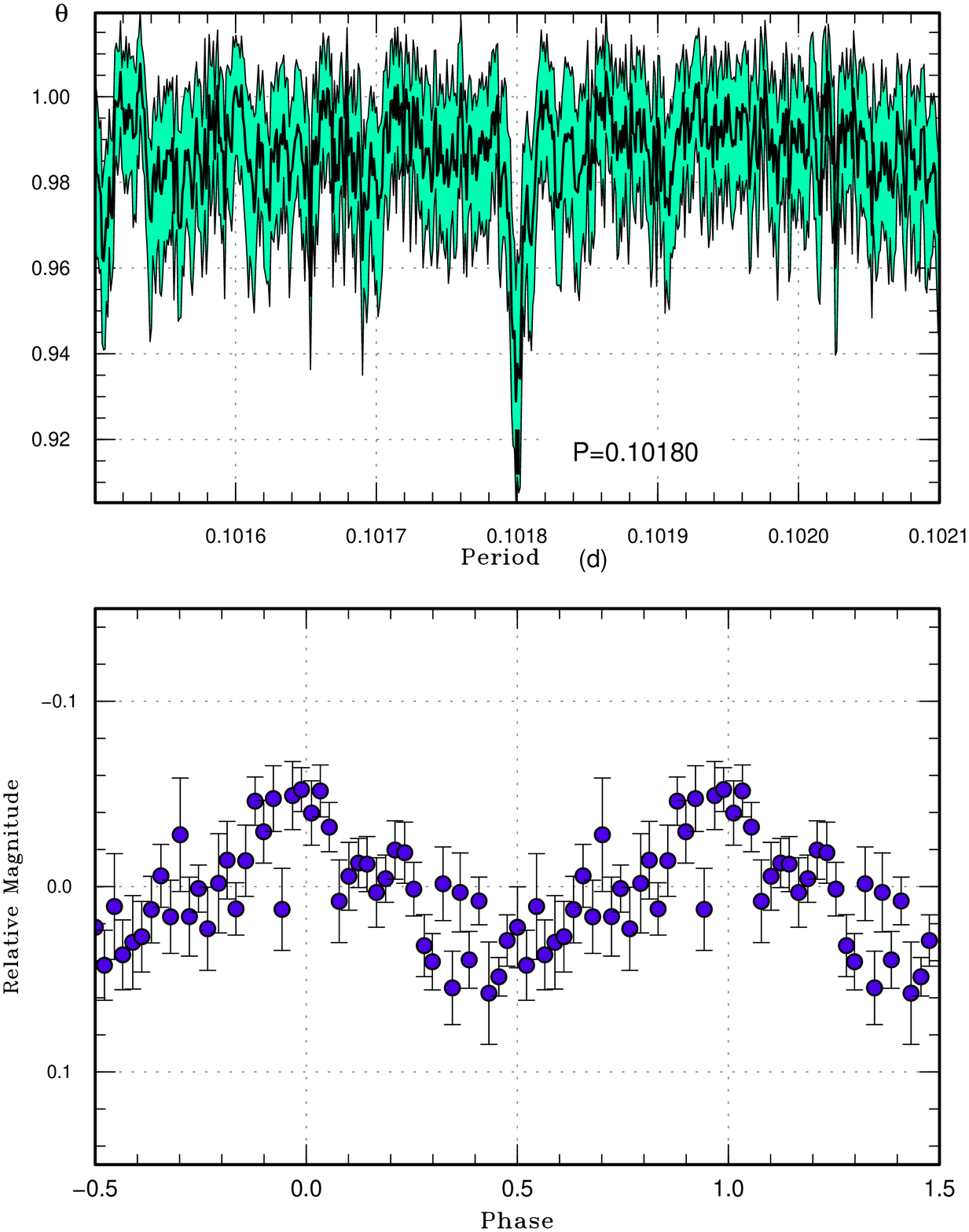}
  \end{center}
  \caption{PDM analysis of V476 Cyg using the ZTF data
  in quiescence.
  (Upper): PDM analysis.  A sharp signal at 0.1018002(6)~d
  was detected.
  (Lower): mean profile.
  }
  \label{fig:v476cygporb}
\end{figure*}

   The mean outburst interval derived from the best
recorded part (BJD 2459300--2459510) was 24.1(1.4)~d.
The durations of most these outbursts were short
(2--3~d), suggesting that V476 Cyg has a relatively
short orbital period.  There was time-resolved photometry
by ZTF on one night (figure \ref{fig:v476cygshort}).
This run suggests a period of $\sim$0.10~d.
With the help of this candidate period, I analyzed
the ZTF data in quiescence (figure \ref{fig:v476cygporb})
using phase dispersion minimization (PDM: \cite{PDM}) analysis
after removing the global trends by locally-weighted polynomial
regression (LOWESS: \cite{LOWESS}).
The error was estimated by methods of \citet{fer89error}
and \citet{Pdot2}.  Although this period looks like
the orbital period, it might come from the physically
unrelated companion star and needs to be confirmed
by further observations.  If this period is the orbital
period of V476 Cyg, this object is in the period gap.
This period appears to be consistent with
the outburst behavior mostly showing short outbursts.
The brightest dwarf nova outburst in the ZTF data had
$M_V$=$+$5.5 using $A_V$=0.7 \citep{sch18novaGaia}
and the Gaia parallax \citep{GaiaEDR3}.  This is
relatively faint among dwarf novae
\citep[see e.g.][]{war87CVabsmag} and appears to be
consistent with a short orbital period.
\citet{kat22wzsgeMV} showed that WZ Sge stars start
showing superhumps at $M_V$=$+$5.4.  The present result
of V476 Cyg is comparable to this value.

   The borders of the period gap is somewhat variable
depending the authors.  I use the range 0.090--0.13~d
based on equation (17) in \citet{kni11CVdonor}.
Well-established novae in the period gap include
IM Nor (recurrent nova) \citep{wou03imnor,pat22imnor},
V Per \citep{sha89vper,sha06vper},
QU Vul \citep{sha95quvul}, V597 Pup \citep{war09v597pup},
and some more borderline or less established cases.
None of these object shows dwarf nova-type outbursts.

\section{Novae showing dwarf nova outbursts after the eruption}

\begin{figure*}
\begin{center}
\includegraphics[width=16cm]{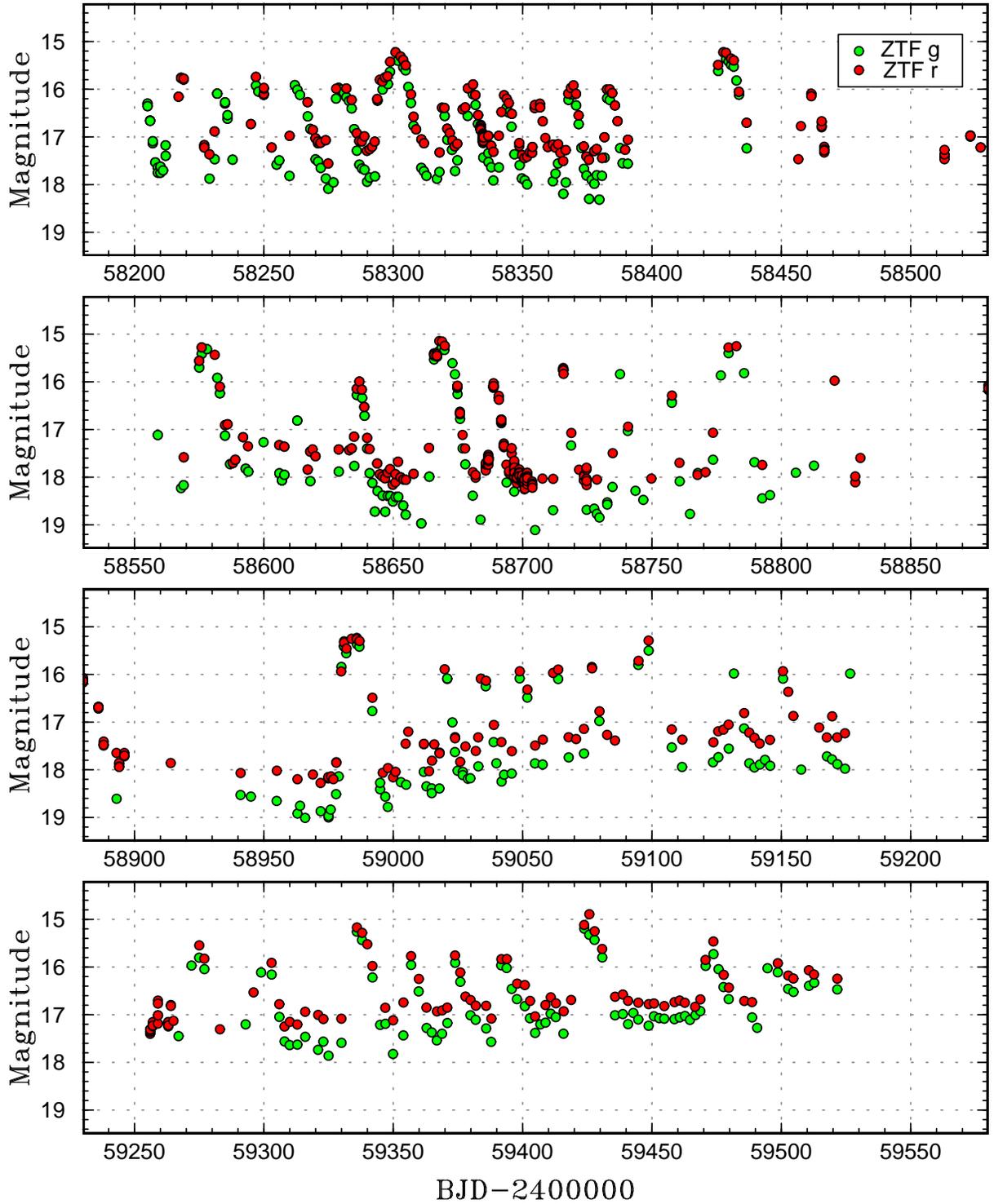}
\caption{
  ZTF light curve of V446 Her.  The current behavior is
  indistinguishable from that of ordinary SS~Cyg-type
  dwarf novae.  The BJD scale is the same as in
  figure \ref{fig:v476cyglc} (V476 Cyg).  One can
  easily see the shortness of outbursts in V476 Cyg.
}
\label{fig:v446herlc}
\end{center}
\end{figure*}

   There are well-established classical novae which currently
show dwarf nova-type outbursts.  I summarized them in
table \ref{tab:novaDN}.  There have been many references
for GK Per and I only listed a few of them.
It might be worth noting that \citet{rob75novapreeruption}
already reported dwarf nova-like outbursts for V446 Her
before the nova eruption.  This phenomenon may have been
similar to the reported case in \citet{mro16hibernation}.
V446 Her currently shows dwarf nova-type outburst typical
for an SS Cyg star with long and short outbursts
in the ZTF data (in contrast to the statement
in \cite{pat13bklyn}: figure \ref{fig:v446herlc}, see also
the light curve in 1994 in \cite{hon95v446her}).
V392 Per was also a dwarf nova
\citep{ric70newvar} before the 2018 nova eruption
\citep[e.g.][]{mun20v392per}, whose most recent slowly rising outburst
was observed in 2016 February--April [detected by
the VSOLJ observer Mitsutaka Hiraga and
the AAVSO observer Carey Chiselbrook (cvnet-outburst message
on 2016 February 28)].
BC Cas is currently in IW And-type state \citep{kat20bccas}
[see e.g. \citet{sim11zcamcamp1,kat19iwandtype} for
IW And-type stars].
A recent light curve for X~Ser is also
present in \citet{kim18j1621}.  A discussion on V1017 Sgr
can be also found in \citet{sal17v1017sgr}.
The most recent dwarf nova-type outburst occurred in 2007.
A outburst of V2109 Oph was detected by
the Gaia satellite as Gaia21dza\footnote{
  $<$http://gsaweb.ast.cam.ac.uk/alerts/alert/Gaia21dza/$>$.
}.  This outburst was a slowly rising one and
the orbital period was suspected to be long
(T. Kato, vsnet-alert 26178\footnote{
  $<$http://ooruri.kusastro.kyoto-u.ac.jp/mailarchive/vsnet-alert/26178$>$.
}).

   BK Lyn was suggested to be the counterpart of
the Chinese ``guest star'' in 101 A.D.
\citep{her86novaID,pat13bklyn}
and showed a transient ER UMa-type phase in
2011--2012 \citep{pat13bklyn,Pdot4,Pdot5}.
This state had apparently started as early as
in 2005 \citep{Pdot4}.  The object is currently in
novalike state and no dwarf nova outbursts are observed.

\begin{table}
\caption{Novae showing dwarf nova outbursts after the eruption}\label{tab:novaDN}
\begin{center}
\begin{tabular}{cccl}
\hline\hline
Object & Eruption & Orbital Period (d) & References \\
\hline
V728 Sco & 1862 & -- & \citet{vog18postnova} \\
V606 Aql & 1899 & -- & \citet{kat21v606aql} \\
GK Per   & 1901 & 1.996803 & \citet{cra86gkperorbit,bia86gkper,sim02gkper} \\
X Ser    & 1903 & 1.478 & \citet{tho00v533herv446herxser,sim18xser} \\
V476 Cyg & 1920 & 0.101800? & this paper \\
BC Cas   & 1929 & -- & \citet{kat20bccas} \\
V446 Her & 1960 & 0.2070 & \citet{tho00v533herv446herxser,hon11v446her} \\
V2109 Oph & 1969 & -- & vsnet-alert 26178 \\
V1017 Sgr & 1919 & 5.78629 & \citet{sek92v1017sgr,web87RN2} \\
\hline
BK Lyn   & 101? & 0.07498 & \citet{rin96bklyn,pat13bklyn} \\
\hline
\end{tabular}
\end{center}
\end{table}

   Although WY Sge (nova eruption in 1783) was once considered to be
a dwarf nova \citep{sha84wysge}, \citet{nay92Hibernation,som96wysge}
pointed out that it is just an ordinary old nova.
Modern ZTF observations do not show any sign of
dwarf nova outbursts contrary to the expectation
by the hibernation scenario \citep{sha84wysge,Hibernation}.
See also \citet{vog18postnova} for modern observations of WY Sge.

   Although \citet{vog18postnova} listed old novae showing
low-amplitude outburst (they referred to as stunted outbursts)
in V841 Oph, V728 Sco, V1059 Sgr, V849 Oph, V363 Sgr,
HS Pup and V2572 Sgr, the dwarf nova-type nature is not
apparent from their light curves for most objects.
I included only V728 Sco, which showed recurrent outbursts
similar to dwarf novae by more than 1~mag, in the table.

   Three of the objects in the table are long-period
systems (orbital periods more than 1~d) and have
evolved secondaries.  It is understandable that
a considerable fraction of this table is composed of
such objects, since these objects have a large accretion disk
and it is unstable to thermal instability even under
mass-transfer rates typical for ordinary (short-period)
novalike systems \citep{kim92gkper}.
\citet{kim92gkper} predicted that outbursts in such
systems are inside-out-type, approximately symmetric ones,
which agree with the observations of these post-novae.
This is apparently not the case for V476 Cyg.
The outbursts in V476 Cyg rise rapidly and
they are apparently outside-in outbursts.
If the suspected orbital period is correct, this behavior
is consistent with the short-period nature.
Among the table, the only confirmed short-period object
is BK Lyn, whose dwarf nova-type phase was likely
a transient phenomenon and the suspected nova eruption
occurred nearly 2000 years ago.  In this regard,
the case of V476 Cyg with a long-lasting dwarf nova-type
phase would be unique.
\citet{pat13bklyn} estimated that novae below
the period gap show dwarf nova outbursts after
the nova eruption when the white dwarf cools sufficiently
after $\sim$1000 years.  If the suspected orbital
period of V476 Cyg is correct, this object can be
an exception.  The case of V476 Cyg may reflect
the rapid evolution (with $t_2$=16.5~d or 6~d) of
the nova eruption and rapid subsequent cooling.

\section{Shoulder or failed superoutburst?}

   The nature of the shoulder in the dwarf nova-type
outburst is not still clear.  \citet{can12ugemLC}
considered it to be similar to precursor outbursts
in SU UMa-type superoutbursts.  \citet{kat21stcha}
suggested that it originates when the disk reaches
the tidal truncation radius.  In the special case
of V363 Lyr, the outburst accompanied by a shoulder
was 0.3--0.4~mag brighter than other outbursts and
showed periodic modulations with a period slightly
longer than the orbital period \citep{kat21v363lyr}.
The nature of this variation is still unclear
\citep{kat21v363lyr}.
Compared to the light curves by \citet{can12ugemLC},
such as that of SS Cyg, the case of V476 Cyg looks more
similar to that of V363 Lyr.  It would be worth
performing time-resolved photometry during such
outbursts to detect possible periodic signals
as in V363 Lyr.  Other shorter outbursts in V476 Cyg
have variable peak brightness, although it was more
constant at 17.0~mag in the late 2020 to the 2021
seasons (later part of the third panel and the fourth
panel of figure \ref{fig:v476cyglc}).
Considering the suspected orbital period in
section \ref{sec:porb}, these outbursts with
shoulders may be analogous to SU UMa-type superoutbursts
[a ``failed superoutburst'' is also known in SU UMa stars,
during which tidal instability is not sufficiently strong
to produce a full superoutburst \citep{osa13v344lyrv1504cyg}],
although the durations were much shorter.

   Determination of the orbital period by radial-velocity
studies is desired.  Considering that many dwarf novae
in the period gap have been identified as SU UMa stars
[e.g. V1006 Cyg \citep{kat16v1006cyg};
MN Dra \citep{nog03var73dra,pav10mndra,bak17mndra};
NY Ser \citep{pav14nyser,kat19nyser}],
superoutbursts may be expected in V476 Cyg.  Continued
observations and timely time-resolved photometry
would clarify the nature of dwarf nova outbursts
in V476 Cyg.  Since the object appears to be still
declining from the 1920 nova eruption, this object
would provide an ideal laboratory of the behavior
of an irradiated accretion disk in which tidal instability
is expected to work.  This object would also be
an ideal laboratory of the effect of a massive white dwarf
on dwarf nova outbursts.

\section*{Acknowledgements}

This work was supported by JSPS KAKENHI Grant Number 21K03616.
The author is grateful to the ZTF team
for making their data available to the public.
We are grateful to Naoto Kojiguchi for
helping downloading the ZTF data.
This research has made use of the AAVSO Variable Star Index
\citep{wat06VSX}, the AAVSO International Database
and NASA's Astrophysics Data System.

Based on observations obtained with the Samuel Oschin 48-inch
Telescope at the Palomar Observatory as part of
the Zwicky Transient Facility project. ZTF is supported by
the National Science Foundation under Grant No. AST-1440341
and a collaboration including Caltech, IPAC, 
the Weizmann Institute for Science, the Oskar Klein Center
at Stockholm University, the University of Maryland,
the University of Washington, Deutsches Elektronen-Synchrotron
and Humboldt University, Los Alamos National Laboratories, 
the TANGO Consortium of Taiwan, the University of 
Wisconsin at Milwaukee, and Lawrence Berkeley National Laboratories.
Operations are conducted by COO, IPAC, and UW.

The ztfquery code was funded by the European Research Council
(ERC) under the European Union's Horizon 2020 research and 
innovation programme (grant agreement n$^{\circ}$759194
-- USNAC, PI: Rigault).

We acknowledge ESA Gaia, DPAC and the Photometric Science
Alerts Team (http://gsaweb.ast.cam.ac.uk/\hspace{0pt}alerts).

\section*{List of objects in this paper}
\xxinput{objlist.inc}

\section*{References}

We provide two forms of the references section (for ADS
and as published) so that the references can be easily
incorporated into ADS.

\renewcommand\refname{\textbf{References (for ADS)}}

\newcommand{\noop}[1]{}\newcommand{\hyphalt}{-}

\xxinput{v476cygaph.bbl}

\renewcommand\refname{\textbf{References (as published)}}

\xxinput{v476cyg.bbl.vsolj}

\end{document}